\documentclass[conference]{IEEEtran}
\usepackage{indentfirst}
\IEEEoverridecommandlockouts
\usepackage{cite}
\usepackage{amsmath,amssymb,amsfonts}
\usepackage{algorithmic}
\usepackage{graphicx}
\usepackage{textcomp}
\usepackage{xcolor}
\usepackage{color,soul}
\usepackage{hyperref}
\usepackage{tabularx}
\usepackage{booktabs}
\usepackage{multirow}

\def\BibTeX{{\rm B\kern-.05em{\sc i\kern-.025em b}\kern-.08em
    T\kern-.1667em\lower.7ex\hbox{E}\kern-.125emX}}
\begin{document}

\title{A journey towards the most efficient state database for Hyperledger Fabric}

\author{\IEEEauthorblockN{Ivan Laishevskiy}
\IEEEauthorblockA{\textit{Idea Blockchain Research Lab} \\
Moscow, Russia \\
ivan.laishevskii@scientificideas.org}
\and
\IEEEauthorblockN{Artem Barger}
\IEEEauthorblockA{\textit{Idea Blockchain Research Lab} \\
Haifa, Israel \\
bartem@scientificideas.org}
\and
\IEEEauthorblockN{Vladimir Gorgadze}
\IEEEauthorblockA{\textit{Moscow Institute of Physics and Technology} \\
Moscow, Russia \\
gorgadze.vv@mipt.ru}
}

\IEEEoverridecommandlockouts
\IEEEpubid{\makebox[\columnwidth]{979-8-3503-1019-1/23/\$31.00~\copyright2023 IEEE \hfill} \hspace{\columnsep}\makebox[\columnwidth]{ }}
\maketitle
\IEEEpubidadjcol

\begin{abstract}
The Hyperledger Fabric is well known and the most prominent enterprise-grade permissioned blockchain. The architecture of the Hyperledger Fabric 
introduces a new architecture paradigm of simulate-order-validate and pluggable architecture, allowing a greater level of customization 
where one of the critical components is the world state database, which is responsible for capturing the snapshot of the blockchain application state. 
Hyperledger Fabric manages the state with the key-value database abstraction and peer updates it after transactions have been validated and read 
from the state during simulation. Therefore, providing good performance during reading and writing impacts the system's overall performance. 
Currently, Hyperledger Fabric supports two different implementations of the state database. One is LevelDB, the embedded DB based on 
LSM trees and CouchDB. In this study, we would like to focus on searching and exploring the alternative implementation of a state database and analyze whenever there are better and more scalable options. We evaluated different databases to be plugged into 
Hyperledger Fabric, such as RocksDB, Boltdb, and BadgerDB. The study describes how to plug new state databases and performance results based on 
various workloads.
\end{abstract}


\section{Introduction}

The increasing popularity of blockchain technology~\cite{l66, l49,l47,l48} is paving the way for businesses to start using blockchain-based solutions particularly those implemented with Hyperledger Fabric~\cite{l50,l46,l67}. Hyperledger Fabric (HLF) is an open-source initiative devoted to creating an enterprise-grade, permissioned blockchain platform~\cite{l62,l63}.

The State Database is one of HLF's most important elements (StateDB). StateDB stores a snapshot of the last blockchain network state. HLF peer relies on the StateDB to read the current values with their version to form the Read-Set for transactions. Then during the commit and validation phase, while executing the multi-value concurrency control~\cite{l65} check, peers read from StateDB to compare Read-Set versions. Finally, once validation is over, the peer commits validated transactions into StateDB to reflect recent changes.~\cite{l64}

According to the~\cite{l43}, the commit phase in HLF dominates the transaction processing time and, as a result, constitutes a bottleneck in terms of performance.
In addition, the study demonstrates that the read operation plays an important role when peers replicate transactions.
Clearly, the interaction with the StateDB directly affects the overall performance of the HLF.
Currently, HLF presents two potential StateDB implementations, one based on GoLevelDB and the other on CouchDB.
Therefore, the Fabric community acknowledged the necessity to develop a superior alternative to StateDB~\cite{l2}.
In this paper, we will analyze various options for StateDB and examine the various implementations of the embedded databases based on LSM-trees or B+ trees~\cite{l1, l44}.

\section{Background}\label{secBackground}


There is a need to establish a starting point for comparison and therefore we implemented a benchmark to identify the maximum potential performance improvement that could be achieved by replacing the current implementations with more performant alternatives. The benchmark implements chaincode FixedAsset with two methods, one which results in interaction between the peer and the StateDB and the other which does not. 

Let's examine the total impact of StateDB on HLF performance using StateDB goleveldb, as it was shown that it outperforms CouchDB~\cite{l68}. For this, a number of performance benchmarks were been run.

\begin{figure}[htbp]
\centerline{\includegraphics[width = 9cm, height = 2.5cm]{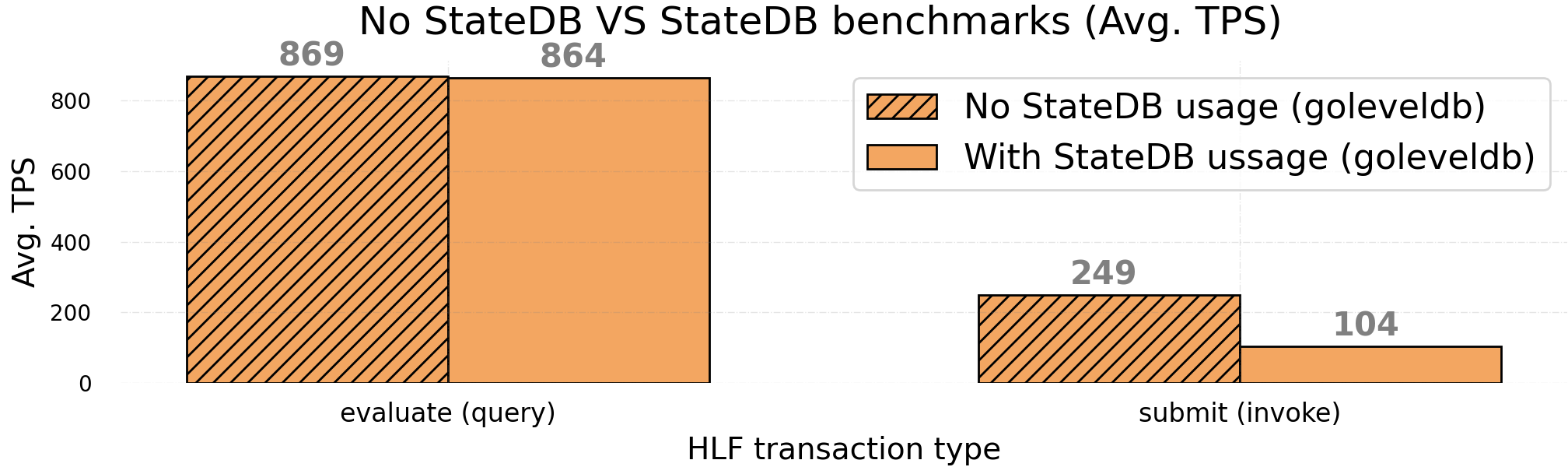}}
\caption{NoStateDB-VS-StateDB performance benchmark results.}
\label{pic6}
\end{figure}

We can conclude that reading 100 byte values in transactions from StateDB has essentially no impact on HLF performance compared to writing the same values to StateDB based on the findings of comparing the average TPS in Fig.\ref{pic6}.

As a result, in this study, we gave top importance to enhancing StateDB's write performance.

\section{Integrating Alternative StateDBs}\label{sec3}

\begin{figure}
  \includegraphics[width=9cm,height=1.3cm]{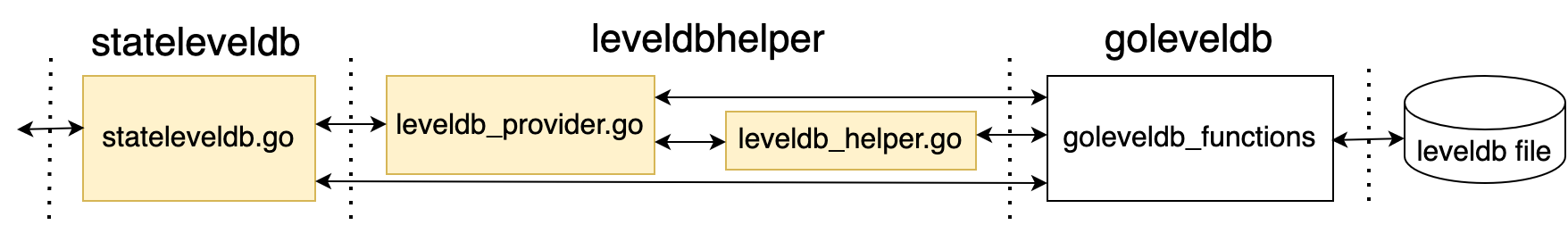}
  \caption{HLF components responsible for interacting with StateDB (highlighted in color).}
\label{pic1}
\end{figure}

In the open source HLF code (commit \cite{l15}), the main components responsible for HLF interaction with StateDB have been identified. These are depicted in Fig. \ref{pic1}. In order to add new StateDBs, it was decided to create an analogue of the marked components for each database. So the main code upgrade was to add three code files for each DB: stateNameOfDB.go, NameOfDB\_provider.go, NameOfDB\_helper.go. Functionality of goleveldb has been fully preserved. To select StateDB you need to specify in core.yaml file the name of selected DB before running the container.

In addition to the modules described, test files covering the added code have also been created and some minor changes have been implemented in some other HLF modules.

\section{Benchmarks for embedded StateDB in HLF}\label{sec4}

\subsection{Performance benchmark evaluation configuration}

Performance evaluations of HLF with the new embedded StateDBs were carried out using the Hyperledger Caliper tool\cite{l35}, which offers universal benchmarks for several blockchain platforms, including Hyperledger Fabric.

This utility supports the capture of basic blockchain performance \cite{l45} metrics. Due to the large number of metrics to compare, in order to simplify for better visualization in our work we will consider only write transactions throughput.

The performance benchmarks are based on code samples from \cite{l36}.

The configuration of the peer machine on which the performance benchmarks were run is: OS: Ubuntu 20.04.4 LTS, CPU(s): 4, RAM: 16, Total SSD memory: 60 GB, Max. bandwidth (read | write): 30 MB/s | 30 MB/s, Max. IOPS (read | write): 2000 | 2000, CPU family: 6, Model: 106, Model name: Intel Xeon Processor (Icelake), Thread(s) per core: 2, Core(s) per socket: 2, Socket(s): 1.

FixedAsset chaincode methods \cite{l38} were developed in Golang to evaluate the performance of various interactions with StateDB.

\subsection{Create-Asset performance benchmark} \label{4c}

The scenario under study is aimed at writing to StateDB. The write performance benchmark consists of Submit calls to the CreateAsset method of the FixedAsset chaincode. We'll refer to such calls as transactions. This chaincode was deployed in independent HLF networks per each considered StateDB: RocksDB, bboltdb or BadgerDB.

Each transaction writes one key-value pair to the StateDB. It lasts 5 minutes for each client for each type of transaction: 100 bytes / 1000 bytes / 4000 bytes / 8000 bytes / 16000 bytes / 24000 bytes / 32000 bytes / 64000 bytes. 

\begin{figure}[htbp]
\centerline{\includegraphics[width = 9cm, height = 2.7cm]{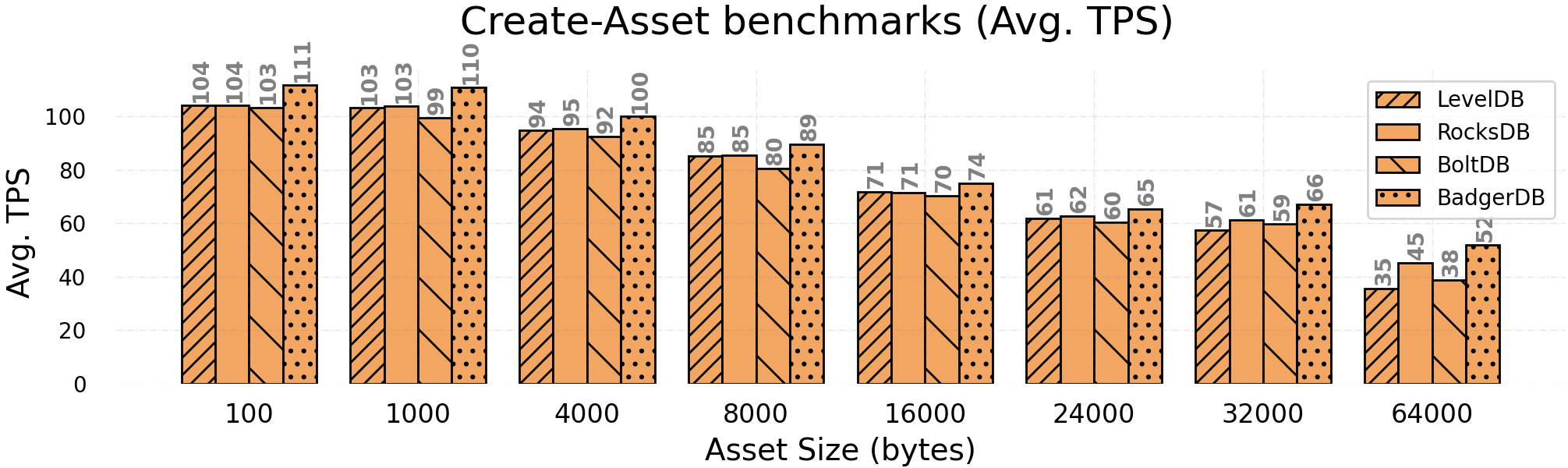}}
\caption{Create-Asset performance benchmark results (transactions per second rate).}
\label{pic305}
\end{figure}

Fig.\ref{pic305} compares the throughput of HLF transactions for embedded StateDBs.
It shows that all StateDB compete with each other almost on an equal footing. However, StateDB BadgerDB has a slight TPS edge for all types of transactions.
Additionally, for a transaction size of 64KB, BadgerDB's TPS is much greater than goleveldb's.
In turn, RocksDB showed equal performance with goleveldb and achieved a clear advantage for a transaction size of 64KB. Similar to this, bbolt only demonstrated a marginal benefit over goleveldb for transactions of 32KB and 64KB in size.

We  can therefore say that BadgerDB demonstrated a higher TPS rate in writing than other StateDBs.


%
%

\section{Conclusions and further direction of work} \label{sec5}

As a result of the research, several databases competing with goleveldb in terms of performance and functionality were selected and embedded in HLF as StateDBs. Furthermore, performance benchmarks were created and conducted for each of the possible alternative StateDBs. In the end, BadgerDB was identified as a favorite of the potential StateDBs.

BadgerDB showed a decent advantage in the StateDB write performance benchmarks. BadgerDB had the strongest advantage over goleveldb for write values of 64KB (TPS is almost 1.5 times higher). In addition, BadgerDB provides StateDB with features that were not implemented with goleveldb: it guarantees ACID properties and makes it possible to use custom queries in HLF by using an additional tool \cite{l29}.

The HLF source code with linked databases is available in the repositories: \cite{l32} (RocksDB), \cite{l33} (bbolt), \cite{l34} (BadgerDB).

To run the performance benchmarks, the HLF fabric-samples/test-network was modified to provide the minimum required set of network components in order to test the performance of StateDB peer with the new databases \url{https://github.com/fubss/fabric-samples/tree/dbs_selector}.

Based on the Hyperledger Caliper \cite{l36} demo, a custom repository was created for the HLF load, adapted for OS Linux and OS X (in different branches) \url{https://github.com/fubss/caliper-workspace-3}.

In further research, it is suggested that:
\begin{itemize}
    \item Compare other performance metrics of embedded databases;
    \item Conduct a read workload performance benchmark;
    \item Identify and eliminate other HLF components that slow it down.
\end{itemize}

\section*{Acknowledgment}

We thank Vladimir Chechetkin for embedding BadgerDB and other contributions to the work on this paper.

\bibliographystyle{IEEEtran}
\bibliography{IEEEabrv,IEEEstatedbrefs.bib}

\begin{thebibliography}{10}
\providecommand{\url}[1]{#1}
\csname url@samestyle\endcsname
\providecommand{\newblock}{\relax}
\providecommand{\bibinfo}[2]{#2}
\providecommand{\BIBentrySTDinterwordspacing}{\spaceskip=0pt\relax}
\providecommand{\BIBentryALTinterwordstretchfactor}{4}
\providecommand{\BIBentryALTinterwordspacing}{\spaceskip=\fontdimen2\font plus
\BIBentryALTinterwordstretchfactor\fontdimen3\font minus
  \fontdimen4\font\relax}
\providecommand{\BIBforeignlanguage}[2]{{%
\expandafter\ifx\csname l@#1\endcsname\relax
\typeout{** WARNING: IEEEtran.bst: No hyphenation pattern has been}%
\typeout{** loaded for the language `#1'. Using the pattern for}%
\typeout{** the default language instead.}%
\else
\language=\csname l@#1\endcsname
\fi
#2}}
\providecommand{\BIBdecl}{\relax}
\BIBdecl

\bibitem{l66}
Y.~Manevich, A.~Barger, and G.~Assa, ``Redacting transactions from
  execute-order-validate blockchains,'' in \emph{2021 IEEE International
  Conference on Blockchain and Cryptocurrency (ICBC)}.\hskip 1em plus 0.5em
  minus 0.4em\relax IEEE, 2021, pp. 1--9.

\bibitem{l49}
J.~Abou~Jaoude and R.~George~Saade, ``Blockchain applications – usage in
  different domains,'' \emph{IEEE Access}, vol.~7, pp. 45\,360--45\,381, 2019.

\bibitem{l47}
S.~M. Nor, M.~Abdul-Majid, and S.~N. Esrati, ``The role of blockchain
  technology in enhancing islamic social finance: the case of zakah management
  in malaysia,'' \emph{foresight}, 2021.

\bibitem{l48}
I.~Fedorov, A.~Pimenov, G.~Panin, and S.~Bezzateev, ``Blockchain in 5g
  networks: Perfomance evaluation of private blockchain,'' in \emph{2021 Wave
  Electronics and its Application in Information and Telecommunication Systems
  (WECONF)}.\hskip 1em plus 0.5em minus 0.4em\relax IEEE, 2021, pp. 1--4.

\bibitem{l50}
\BIBentryALTinterwordspacing
{Hyperledger Foundation}, \emph{A Blockchain Platform for the Enterprise},
  2022, revision dd63f081. [Online]. Available:
  \url{https://hyperledger-fabric.readthedocs.io/en/latest/}
\BIBentrySTDinterwordspacing

\bibitem{l46}
\BIBentryALTinterwordspacing
------. (2022) Case studies. [Online]. Available:
  \url{https://www.hyperledger.org/learn/case-studies}
\BIBentrySTDinterwordspacing

\bibitem{l67}
A.~Barger, O.~Ilina, A.~Zemtsov, and K.~Tagirova, ``Trustful charity foundation
  platform based on hyperledger fabric,'' in \emph{2022 IEEE International
  Conference on Omni-layer Intelligent Systems (COINS)}.\hskip 1em plus 0.5em
  minus 0.4em\relax IEEE, 2022, pp. 1--6.

\bibitem{l62}
\BIBentryALTinterwordspacing
{Hyperledger Foundation}. (2022) Hyperledger fabric. [Online]. Available:
  \url{https://www.hyperledger.org/use/fabric}
\BIBentrySTDinterwordspacing

\bibitem{l63}
E.~Androulaki, A.~Barger, V.~Bortnikov, C.~Cachin, K.~Christidis, A.~De~Caro,
  D.~Enyeart, C.~Ferris, G.~Laventman, Y.~Manevich \emph{et~al.}, ``Hyperledger
  fabric: a distributed operating system for permissioned blockchains,'' in
  \emph{Proceedings of the thirteenth EuroSys conference}, 2018, pp. 1--15.

\bibitem{l64}
A.~Barger, Y.~Manevich, H.~Meir, and Y.~Tock, ``A byzantine fault-tolerant
  consensus library for hyperledger fabric,'' in \emph{2021 IEEE International
  Conference on Blockchain and Cryptocurrency (ICBC)}.\hskip 1em plus 0.5em
  minus 0.4em\relax IEEE, 2021, pp. 1--9.

\bibitem{l65}
C.~H. Papadimitriou and P.~C. Kanellakis, ``On concurrency control by multiple
  versions,'' \emph{ACM Transactions on Database Systems (TODS)}, vol.~9,
  no.~1, pp. 89--99, 1984.

\bibitem{l43}
T.~Nakaike, Q.~Zhang, Y.~Ueda, T.~Inagaki, and M.~Ohara, ``Hyperledger fabric
  performance characterization and optimization using goleveldb benchmark,'' in
  \emph{2020 IEEE International Conference on Blockchain and Cryptocurrency
  (ICBC)}.\hskip 1em plus 0.5em minus 0.4em\relax IEEE, 2020, pp. 1--9.

\bibitem{l2}
\BIBentryALTinterwordspacing
(2022) Leveldb key/value database in go. Github repository. [Online].
  Available: \url{https://github.com/syndtr/goleveldb}
\BIBentrySTDinterwordspacing

\bibitem{l1}
\BIBentryALTinterwordspacing
D.~Enyeart. (2021) Fabric strategic priorities. [Online]. Available:
  \url{https://lists.hyperledger.org/g/fabric/message/10357}
\BIBentrySTDinterwordspacing

\bibitem{l44}
\BIBentryALTinterwordspacing
{Hyperledger Foundation}. (2021) Fabric strategic priorities - 2021 discussion.
  [Online]. Available:
  \url{https://wiki.hyperledger.org/display/fabric/Fabric+Strategic+Priorities+-+2021+discussion}
\BIBentrySTDinterwordspacing

\bibitem{l68}
P.~Thakkar, S.~Nathan, and B.~Viswanathan, ``Performance benchmarking and
  optimizing hyperledger fabric blockchain platform,'' in \emph{2018 IEEE 26th
  international symposium on modeling, analysis, and simulation of computer and
  telecommunication systems (MASCOTS)}.\hskip 1em plus 0.5em minus 0.4em\relax
  IEEE, 2018, pp. 264--276.







\bibitem{l15}
\BIBentryALTinterwordspacing
(2021) Evaluate() error response for node chaincode. Commit in Github
  repository. [Online]. Available:
  \url{https://github.com/hyperledger/fabric/commit/6656f72563c73a806dee7068dd91b3acf2a286aa}
\BIBentrySTDinterwordspacing


\bibitem{l32}
\BIBentryALTinterwordspacing
(2022) Hyperledger fabric with embedded grocksdb. Branch in Github repository.
  [Online]. Available: \url{https://github.com/fubss/fabric/tree/grocksdb-30}
\BIBentrySTDinterwordspacing

\bibitem{l33}
\BIBentryALTinterwordspacing
(2022) Hyperledger fabric with embedded bbolt. Branch in Github repository.
  [Online]. Available: \url{https://github.com/fubss/fabric/tree/bbolt-30}
\BIBentrySTDinterwordspacing

\bibitem{l34}
\BIBentryALTinterwordspacing
(2022) Hyperledger fabric with embedded badgerdb. Branch in Github repository.
  [Online]. Available: \url{https://github.com/fubss/fabric/tree/badger-30}
\BIBentrySTDinterwordspacing

\bibitem{l35}
\BIBentryALTinterwordspacing
(2022) Hyperledger caliper. Github repository. [Online]. Available:
  \url{https://github.com/hyperledger/caliper}
\BIBentrySTDinterwordspacing

\bibitem{l45}
\BIBentryALTinterwordspacing
{Performance and Scale Working Group (PSWG)}. (2018) Definitions of key
  metrics. Article. [Online]. Available:
  \url{https://www.hyperledger.org/learn/publications/blockchain-performance-metrics#definitions}
\BIBentrySTDinterwordspacing

\bibitem{l36}
\BIBentryALTinterwordspacing
(2022) Hyperledger caliper benchmarks. Github repository. [Online]. Available:
  \url{https://github.com/hyperledger/caliper-benchmarks}
\BIBentrySTDinterwordspacing

\bibitem{l38}
\BIBentryALTinterwordspacing
(2022) Fixedassetcontract. File in Github repository. [Online]. Available:
  \url{https://github.com/fubss/caliper-workspace-3/blob/main/smart-contract/go/FixedAssetContract.go}
\BIBentrySTDinterwordspacing



\bibitem{l29}
\BIBentryALTinterwordspacing
(2022) Badgerhold. Github repository. [Online]. Available:
  \url{https://github.com/timshannon/badgerhold}
\BIBentrySTDinterwordspacing

\end{thebibliography}
\end{document}